\documentclass[conference]{IEEEtran}
\ifCLASSINFOpdf
\else
\fi
\usepackage{dsfont}
\usepackage{bm}
\usepackage{makecell}
\usepackage{amsmath}
\usepackage{amssymb}
\usepackage{algorithm}
\usepackage{algorithmic}
\usepackage{multirow}
\usepackage{xcolor}
\usepackage{graphics}
\usepackage{graphicx}
\usepackage{epsfig}
\usepackage{enumerate}

\newtheorem{theorem}{Theorem}
\newtheorem{definition}{Definition}

\newtheorem{lemma}{Lemma}

\begin{document}
%
% paper title
% Titles are generally capitalized except for words such as a, an, and, as,
% at, but, by, for, in, nor, of, on, or, the, to and up, which are usually
% not capitalized unless they are the first or last word of the title.
% Linebreaks \\ can be used within to get better formatting as desired.
% Do not put math or special symbols in the title.
\title{Release Connection Fingerprints in Social Networks Using Personalized Differential Privacy}

% author names and affiliations
% use a multiple column layout for up to three different
% affiliations
\author{\IEEEauthorblockN{Yongkai Li\IEEEauthorrefmark{1}\IEEEauthorrefmark{2},
Shubo Liu\IEEEauthorrefmark{1}\IEEEauthorrefmark{2},
Jun Wang\IEEEauthorrefmark{1}\IEEEauthorrefmark{2},
and Mengjun Liu\IEEEauthorrefmark{1}\IEEEauthorrefmark{2}}
\IEEEauthorblockA{\IEEEauthorrefmark{1}School of Computer, Wuhan University, Wuhan, China}
\IEEEauthorblockA{\IEEEauthorrefmark{2}Key Laboratory of Aerospace Information Security and Trusted Computing, Ministry of Education,\\Wuhan University, Wuhan, China \\
Email: whu\_lyk@foxmail.com,  liu.shubo@whu.edu.cn, jameswang@whu.edu.cn,  mengjun@hawaii.edu}}
\maketitle

% As a general rule, do not put math, special symbols or citations
% in the abstract
\begin{abstract}
There are many benefits of publication social networks statistics for societal or commercial purposes, such as political advocacy and product recommendation. It is very challenging to protect the privacy of individuals in social networks while ensuring a high accuracy of the statistics. Moreover, most of the existing work on differentially private social network publication ignores the facts that different users may have different privacy preferences and there also exists a considerable amount of users whose identities are public. In this paper, we aim to release the number of public users that a private user connects to within $n$ hops (denoted as $n$-range Connection Fingerprints,or $n$-range CFPs for short) regarding user-level personalized privacy preferences. To this end, we proposed two schemes, DEBA and DUBA-LF, for privacy-preserving publication of the CFPs on the base of personalized differential privacy(PDP), and conduct a theoretical analysis of the privacy guarantees provided within the proposed schemes. The implementation showed that the proposed schemes are superior in publication errors on real datasets.
\end{abstract}

% no keywords

% For peer review papers, you can put extra information on the cover
% page as needed:
% \ifCLASSOPTIONpeerreview
% \begin{center} \bfseries EDICS Category: 3-BBND \end{center}
% \fi
%
% For peerreview papers, this IEEEtran command inserts a page break and
% creates the second title. It will be ignored for other modes.
\IEEEpeerreviewmaketitle

\section{Introduction}
Nowadays, more and more people join multiple social networks on the Web, such as Facebook, Twitter, and Sina Weibo, to share their own information and at the same time to monitor or participate in different activities. Many institutions and firms are investing time and resources into analyzing social networks to address a diverse set of societally or commercially important issues including disease transmission, product recommendation, and political advocacy, among many others. Although the sophistication of information technology has allowed the collection, analysis, and dissemination of social network data, privacy concerns have posed significant restriction of the ability of social scientists and others to study these networks. To respect the privacy of individual participants in the networks, network data cannot be released for public access and scientific studies without proper sanitization.

A common practice is to release a ``naively anonymized'' isomorphic network after removing the real identities of vertices. It is now well-known that this can leave participants open to a range of attacks \cite{1,2}. Thus, a great many of anonymization techniques have been proposed \cite{2,3,4,5,6} to ensure network data privacy. However, those anonymization techniques have been shown to be susceptible to several newly discovered privacy attacks and lack of rigorous privacy and utility guarantees. In response, differential privacy \cite{7} has been applied to solve such vulnerability in social network data publication. Differential privacy is a popular statistical model, and it prevents any adversary from inferring individual information from the output of a computation by perturbing the data prior to the release. A limitation of the model is that the same level of privacy protection is afforded for all individuals. However, it is common that different users may have \emph{different privacy preferences}\cite{18,34,8}. Therefore, providing the same level privacy protection to all the users may not be fair and in addition may cause the published social network data useless. Moreover, in reality, not all the identities of social network users are sensitive\cite{6}. For instance, Sina Weibo, a popular Chinese microblogging social network, hosts around lots of media accounts (e.g., NBA and Xinhuanet), and millions of celebrity accounts (e.g., Christine Lagarde and Kai-Fu Lee). All these users¡¯ identities are public, and they in total account for over 1\% of the overall half billion registered user accounts \cite{9}.

In this paper, we classify the users whose identities are not sensitive as \emph{public users} to be distinguished from \emph{private users}. It is pointed out that releasing the identities of public users with social network data can benefit both research and the users themselves \cite{6}. Moreover, we take different privacy requirements into account to guarantee precisely the required level of privacy to different users.

In this work, we focus on a specific publication goal when public users are labeled, i.e., the number of public users that a private user connects to within $n$ hops. For ease of presentation, we also use the definition \emph{n-Range connection fingerprints} (CFPs)\cite{6} to denote the public users that a private user connects to within $n$ hops. We choose to focus on the number of CFPs because it is one of the most important properties for a public users labeled graph. For example, these statistics can be used for studying the social influence of government organizations, simulating information propagation through media, helping corporate make smart targeted advertising plans, and so on.

In this work, we consider the setting in which a trusted data analyst desires to publish the number of $n$-Range CFPs of each private user. Every private user potentially requires different privacy guarantee for his or her statistics and the analyst would like to publish useful aggregate information about the network. To this end, we employ a new privacy framework, Personalized Differential Privacy (PDP)\cite{34,8}, to provide personal privacy guarantee specified at the user-level, rather than by a single, global privacy parameter. The privacy guarantees of PDP have the same strength and attack resistance as differential privacy, but are personalized to the preferences of all users in the input domain.

In this work, we propose two schemes to release the number of CFPs in the context of personalized privacy preferences. We address the challenge of improving the data utility by employing the distance-based approximation mechanism and decreasing the introduced noise. The main contributions of this paper are:
\begin{enumerate}
  \item To the best our knowledge, we formalized the question of releasing the number of CFPs in the context of personalized privacy preferences for the first time.
  \item We present two schemes, DEBA and DUBA-LF, for privacy-preserving publication of the CFPs regarding personalized privacy preferences, and we conduct a theoretical analysis of the privacy guarantees provided within the proposed schemes. The proposed schemes are designed to be $\mathcal{P}$-PDP.
\item We experimentally evaluate the two proposed schemes on real datasets and it is demonstrated that our proposed schemes have high utility for each dataset.
\end{enumerate}

The paper is organized as follows. Section \uppercase\expandafter{\romannumeral2} discusses preliminaries and related work. Section \uppercase\expandafter{\romannumeral3} presents the problem statement and privacy goal. Overview of our solutions is described in Section \uppercase\expandafter{\romannumeral4}. Section \uppercase\expandafter{\romannumeral5} presents our methods for privacy-preserving CFPs publishing. The privacy analysis is reported in Section \uppercase\expandafter{\romannumeral6}. Section \uppercase\expandafter{\romannumeral7} describes some of our experimental results and performance analysis. Section \uppercase\expandafter{\romannumeral8} presents the conclusions of this research.

\section{Preliminaries}
In this section, we introduce some notations and initial definitions, and review the definition of differential privacy, two conventional mechanisms to achieve differential privacy, upon which our work is based. Then the related work is discussed.

We model a social network as an undirected and unweighted graph $G =(V,E) \in \mathcal{G}$, where $V$ is a set of vertices representing user entities in the social network, and $E$ is a set of edges representing social connections between users (e.g., friendships, contacts, and collaborations). The notation $e(v_i , v_j) \in E$ represents an edge between two vertices $v_i$ and $v_j$ . We let $|V|$ = $n_0$ and the notation $|V|$ is used to represent the cardinality of $V$. For ease of presentation, we use ``grap'' and ``social network'' interchangeably in the following discussion, as well as ``user'' and ``node''.

\subsection{Differential Privacy}
We call two graphs $G$, $G'$ as neighboring if $G'$ can be obtained from $G$ by removing or adding a one edge, i.e., their minimum edit distance \cite{10} $d(G, G') \leq 1$. We write $G \xrightarrow{e} G'$  to denote that $G$ and $G'$ are neighbors and that $G = G' \wedge e$ or $G' = G\wedge e$, where $e$ is an egde. Differential privacy requires that, prior to $f(G)$'s release, it should be modified using a randomized algorithm $\mathcal{A}$, such that the output of $\mathcal{A}$ does not reveal much information about any edge in $G$. The definition of differential privacy is shown as follows:
\begin{definition}
($\epsilon$-differential privacy)\cite{7}. A randomized algorithm $\mathcal{A}$ is $\epsilon$-differentially private if for any two graphs $G$ and $G'$  that are neighboring, and for all $O \in Range(\mathcal{A})$,
$Pr[\mathcal{A}(G) \in O] \leq e^{\epsilon} \cdot Pr[\mathcal{A}(G') \in O]$.
\end{definition}

A differentially private algorithm $\mathcal{A}$ provides privacy because, given any two graphs which differ on a single edge only, respective results of a same query on the graphs are not distinguishable. Therefore, an adversary cannot infer the value of any single edge in the dataset. Here, $\epsilon$ represents the level of privacy. A smaller value of $\epsilon$  means better privacy, but it also implies lower accuracy of the query result. The composition of differentially private algorithms also provides differential privacy, but it produces different results depending on the data to which the queries are applied.
\begin{itemize}
\item \textbf{Sequential composition [\cite{11}, Theorem 3]}. Let $\mathcal{A}_i$ each provides $\epsilon_i$-differential privacy. The sequence of $\mathcal{A}_i(X)$ provides $(\sum_{i}\epsilon_i)$-differential privacy.
\end{itemize}

While there are many approaches to achieving differential privacy, the best known and most-widely used two for this purpose are the Laplace mechanism \cite{12} and the exponential mechanism \cite{13}. For real valued functions, i.e., $f:\mathcal{G} \rightarrow R^d$, the most common way to satisfy differential privacy is to inject carefully chosen random noise into the output. The magnitude of the noise is adjusted according to the global sensitivity of the function, or the maximum extent to which any one tuple in the input can affect the output. Formally,
\begin{definition}
(Global Sensitivity \cite{12}): The global sensitivity of the function $f:\mathcal{G} \rightarrow R^d$ is $\Delta(f)= \max_{d(G,G')\leq 1} \|f(G)-f(G')\|$ for all neighboring $G, G' \in \mathcal{G}$, where $\|\cdot\|$ denotes the $L_{1}$ norm.
\end{definition}

Similarly, the local sensitivity and local sensitivity at distance $t$ of function $f$ are defined as follows.
\begin{definition}
(Local Sensitivity \cite{14}): The local sensitivity of the function $f:\mathcal{G} \rightarrow R^d$ is $LS(G,f)= \max_{G'|d(G,G')\leq 1} \|f(G)-f(G')\|$ , where $\|\cdot\|$ denotes the $L_{1}$ norm.
\end{definition}
\begin{definition}
(Local Sensitivity at distance $t$ \cite{14}): The local sensitivity of $f$ at distance $t$ is the largest local sensitivity attained on graphs at distance at most $t$ from $G$. Formally, the global sensitivity of the function $f:\mathcal{G} \rightarrow R^d$ at distance $t$ is $LS(G,f,t)= \max_{G'|d(G,G')\leq t} \|f(G)-f(G')\|$ , where $\|\cdot\|$ denotes the $L_{1}$ norm.
\end{definition}

Note that global sensitivity can be understood as the maximum of local sensitivity over the input domain, i.e., $\Delta(f) = \max_{G} LS(G, f)$ and local sensitivity of $f$ is a special case of $LS(G, f , t)$ for distance $t=1$.

To maintain differential privacy, the Laplace mechanism adds noise drawn from the Laplace distribution into the data to be published. The influence of any single edge on the outcome will be masked and hidden by the Laplace noise. Let $Lap(\lambda)$ be a random value sampled from a Laplace distribution with mean zero and scale $\lambda$. The Laplace Mechanism through which $\epsilon$-differential privacy is achieved is outlined in the following theorem.
\begin{theorem}
\cite{11}Let $f:\mathcal{G} \rightarrow R^d$. A mechanism $M$ that adds independently generated noise from a zero-mean Laplace distribution with scale $\lambda=\Delta(f)/\epsilon$ to each of the $d$ output values $f(G)$, i.e., which produces $O = f(G)+ \langle Lap(\Delta(f)/\epsilon)\rangle^d$ satisfies $\epsilon$-differential privacy.
\end{theorem}

The exponential mechanism \cite{13} is useful for sampling one of several options in a differentially-private way. A score to each of the options, which is determined by the input of the algorithm, is assigned by a quality function $q$. Clearly, higher scores signify more desirable outcomes and the scores are then used to formulate a probability distribution over the outcomes in a way that ensures differential privacy.
\begin{definition}
(Exponential Mechanism \cite{12}).Let $q:(\mathcal{G}\times\mathcal{O}) \rightarrow R$ be a quality function that assigns a score to each outcome $O\in \mathcal{O}$. Let $\Delta_1(q)= max_{d(G,G')\leq 1}\left \|q(G,O)-q(G',O)\right \|$ and $M$ be a mechanism for choosing an outcome $O\in \mathcal{O}$. Then the mechanism $M$, defined by \\ $M(G,q)=\{$return $O$ with probability $\propto exp(\frac{\epsilon q(G,O)}{2\Delta_1(q)})\}$\\  maintains $\epsilon$-differential privacy.
\end{definition}

\subsection{Related Work}
With the increasing popularity of social network analysis research, privacy protection of social network data is a broad topic with a significant amount of prior work. In this section, we review the most closely related work about privacy protection on social network data.

An important thread of research aims to preserve social network data privacy by obfuscating the edges (vertices), i.e., by adding /deleting edges (vertices).\cite{6,15,16,17}.  Mittal et al. proposed a perturbation method in \cite{15} by deleting all edges in the original graph and replacing each edge with a fake edge that is sampled based on the structural properties of the graph. Liu et al. \cite{16} design a system, called LinkMirage , which mediates privacy-preserving access to users¡¯ social relationships in both static and dynamic social graphs. Hay et al. \cite{17} perturb the graph by applying a sequence of $r$ edge deletions and $r$ edge insertions. The deleted edges are uniformly selected from the existing edges in the original graph while the added edges are uniformly selected from the non-existing edges.
Wang et al. propose two different perturbation methods to anonymize social networks against CFP attacks in \cite{6}, which serves as the practical foundation for our algorithm. Their first method is based on adding dummy vertices, while the second algorithm achieves $k$-anonymity based on edge modification. Their proposed methods can resist CFP attacks on private users based on their connection information with the public users. Another important work for our algorithm is \cite{18}. Yuan et al. \cite{18} introduce a framework that provides personalized privacy protection for labeled social networks. They define three levels of privacy protection requirements by modeling gradually increasing adversarial background knowledge. The framework combines label generalization and other structure protection techniques (e.g., adding nodes or edges) in order to achieve improved utility.

Most of the obfuscating based works mainly focus on developing anonymization techniques for specific types of privacy attacks. They employ privacy models derived from $k$-anonymity \cite{19} by assuming different types of adversarial knowledge. Unfortunately, all these anonymization techniques are vulnerable to attackers with stronger background knowledge than assumed, which has stimulated the use of differential privacy for more rigorous privacy guarantees.

There are many papers \cite{20,21,22,23,24,25} have started to apply differential privacy to protect edge/node privacy, defined as the privacy of users¡¯ relationship or identity in graph data. One important application direction aims to release certain differentially private data mining results, such as degree distributions, subgraph counts and frequent graph patterns \cite{21,23,25}. However, our problem is substantially more challenging than publishing certain network statistics or data mining results. Our goal is to publish the CFPs of private user, which incurs a much larger global sensitivity. Note that the sensitivity in the problem setting of \cite{23} is only 1. What¡¯s more, each private user in the networks independently specifies the privacy requirement for their data. In addition, some latest works are done for graph-oriented scenario. Proserpio et al. \cite{24} develop a private data analysis platform wPINQ over weighted datasets, which can be used to answer subgraph-counting queries. Zhang et al. \cite{25} propose a ladder framework to privately count the number of occurrences of subgraphs.

There are also some other related works aiming to publish a sanitized graph, which is out the scope of the objective of this paper. In \cite{26}, Sala et al. introduced Pygmalion, a differentially private graph model. Similar to \cite{26}, Wang and Wu employed the dK-graph generation model for enforcing edge differential privacy in graph anonymization \cite{27}. In \cite{28}, Xiao et al. proposed a Hierarchical Random Graph (HRG) model based scheme to meet edge differential privacy. In addition, Chen et al. \cite{29} propose a data-dependent solution by identifying and reconstructing the dense regions of a graph's adjacency matrix.

\section{Privacy Goal}

\subsection{Problem Definition}
In general, we assume there are some public users whose identities are not sensitive in social networks. Except the public users, the rest of the users are private users, whose edges are sensitive and each private user independently specifies the privacy requirement for their data. For convenience, we denote the set of public users in the social network as $V_{pub}$ and the set of private users $V_{pri}$ and we let  $|V_{pub}|= m_p $ and $|V_{pri}|= m $. More formally, the Privacy Specification of private users is defined as follows:
\begin{definition}
(User-Level Privacy Specification\cite{8}). A privacy specification is a mapping $\mathcal{P}: V_{pri} \rightarrow R^+$ from private users to personal privacy preferences, where a smaller value represents a stronger privacy preference. The notation $P^v$ is used to denote the privacy preference corresponding to user $v \in V_{pri}$.
\end{definition}

Similar to \cite{8}, we also describe a specific instance of a privacy specification as a set of ordered pairs, e.g., $\mathcal{P}:= \{(v_1, \epsilon_1),(v_2, \epsilon_2),\ldots\}$ where $v_i \in V_{pri}$ and $\epsilon_i\in R^+$. We also assume that a privacy specification contains a privacy preference for every $v \in V_{pri}$ , or that a default privacy level is used. Here the information about any edge in $G$ should be protected, and the privacy specification of edge $e(v_i,v_j)$ can be quantified by $min\{P^{v_i},P^{v_j}\}$.

In this paper, we focus on privately releasing of connection statistics between private users and public users. First, we specify the hop distance $h(v_i,v_j)$ between two vertices $v_i$ and $v_j$ as the number of edges on the shortest path between them. Second, as indicated in ref.\cite{6}, we call the public user $v$ as $v_i$'s \emph{connection fingerprint (CFP)} if the private user $v_i$ and the public user $v$ is linked by a path of certain length. For a given hop distance $n$, the formal definitions of the $n$th-hop connection fingerprint $CFP_n(v_i)$ and $n$-range connection fingerprint $CFP(v_i, n)$ for a private user $v_i$ in Definition 7 and Definition 8, respectively.
\begin{definition}
($n$th-Hop Connection Fingerprint\cite{6}): The $n$th-hop connection fingerprint $CFP_n(v_i)$ of a private vertex $v_i$ in a social network $G=(V,E)$ consists of the group of public vertices whose hop distances to $v_i$ are exactly $n$, i.e., $CFP_n(v_i)= \{ID(v_j)| v_j \in V_{pub}\wedge h(v_i,v_j)= n\}$.
\end{definition}

\begin{definition}
($n$-Range Connection Fingerprint\cite{6}): The $n$-range connection fingerprint of a private vertex $v_i$, denoted by $CFP(v_i, n)$, is formed by $v_i$'s $x$th-hop connection fingerprints, where $1 \leq x \leq n$, i.e., $CFP(v_i, n) = \cup_{x \in [1,n]}{CFP_x(v_i)}$.
\end{definition}

Given a system parameter $c$, we aim to release the number of $k$th-hop($1 \leq k \leq c$) connection fingerprints for each private user in the sensitive graph, while protecting individual privacy in the meantime. Formally, we write $f_k(G):\mathcal{G} \rightarrow R^m$ to denote the function that computes the number of $k$th-hop connection fingerprints for each private user in graph $G$. Therefore, the final publication results for a sensitive graph $G$ can be denoted in a form of $m \times c$ matrix $F=(f_1(G),¡­¡­, f_c(G))$.

\subsection{Privacy Goal}
The goal of this paper is to release the connection statistics under the novel notation of Personalized Differential Privacy (PDP) \cite{34,8}. In contrast to traditional differential privacy, in which the privacy guarantee is controlled by a single, global privacy parameter (i.e., $\epsilon$), PDP makes use of a privacy specification, in which each user in $V_{pri}$ independently specifies the privacy requirement for their data. More formally, the definition of PDP is showed in Definition 9.
\begin{definition}
(Personalized Differential Privacy (PDP)\cite{34,8}). In the context of a privacy specification $\mathcal{P}$ and a universe of private users $U$, a randomized mechanism $\mathcal{M}:\mathcal{G} \rightarrow R^m$ satisfies $\mathcal{P}$-personalized differential privacy (or $\mathcal{P}$-PDP), if for every pair of neighboring graphs $G$ and $G'$, with $G \xrightarrow{e_{ij}} G'$   and $e_{ij} = e (v_i , v_j)$, and for all $O \in Range(\mathcal{M})$,
$Pr[\mathcal{M}(G) \in O] \leq e^{min\{P^{v_i},P^{v_j}\}} \cdot Pr[\mathcal{M}(G') \in O]$.
\end{definition}

Intuitively, PDP offers the same strong, semantic notion of privacy that traditional differential privacy provides, but the privacy guarantee for PDP is personalized to the needs of every user simultaneously. Jorgensen et al.\cite{8} point out that the composition properties of traditional differential privacy extend naturally to PDP, see Theorem 2.
\begin{theorem}
(Composition\cite{8}). Let $\mathcal{M}_1:\mathcal{G} \rightarrow R^m$ and $\mathcal{M}_2:\mathcal{G} \rightarrow R^m$ denote two mechanisms that satisfy PDP for $\mathcal{P}_1$ and $\mathcal{P}_2$ , respectively. Then, the mechanism $\mathcal{M}_3: = (\mathcal{M}_1(\mathcal{G}), \mathcal{M}_2(\mathcal{G}))$ satisfies $\mathcal{P}_3$-PDP, where $\mathcal{P}_3= \mathcal{P}_1 + \mathcal{P}_2$.
\end{theorem}

A smart general purpose mechanism for achieving PDP, called Sample Mechanism, is proposed in \cite{8}. The sample mechanism works by introducing two independent sources of randomness into a computation: (1) non-uniform random sampling at the tuple level, and (2) additional uniform randomness introduced by invoking a traditional differentially private mechanism on the sampled input.
\begin{theorem}
(The Sample Mechanism\cite{8}). Consider a function $f:\mathcal{G} \rightarrow R^m$, a social network $G \in \mathcal{G}$, a configurable threshold $t$ and a privacy specification $\mathcal{P}$. Let $RS(G, \mathcal{P}, t)$ denote the procedure that independently samples each edge $e_{ij} = e(v_i , v_j) \in G$ with probability
\\
\[\pi(e_{ij},t) = \left\{\begin{array}{ll}
\frac{e^{\min\{P^{v_i},P^{v_j}\}}-1}{e^t-1}&\text{if $\min\{P^{v_i},P^{v_j}\}<t$},\\
1&\text{otherwise}.
\end{array}\right.\]
where $\min_v P^v \leq t \leq \max_v P^v$. The sample mechanism is defined as $S_f (G, \mathcal{P}, t) = DP_{f}^{t}(RS(G, \mathcal{P}, t))$ where $DP_{f}^{t}$ is any $t$-differentially private mechanism that computes the function $f$. Then the sample mechanism $S_f (G, \mathcal{P}, t)$ achieves $\mathcal{P}$-PDP.

The mechanism $DP_{f}^{t}$ could be a simple instantiation of the Laplace or exponential mechanisms, or a more complex composition of several differentially private mechanisms.
\end{theorem}

\section{Overview of our solutions}
Given our problem of releasing the number of $k$th-hop($1 \leq k \leq c$) connection fingerprints under $\mathcal{P}$-PDP, we overview the baseline and our advanced methods to give a brief glance on our motivations.

\textbf{Baseline method.} A baseline method is to apply a(n) uniform or exponential budget allocation method and release a $\mathcal{P}/c$ (or $\mathcal{P}/2^k$ and the budget is $\mathcal{P}/2^{c-1}$ indeed for $k = c$)-personalized differential private result for every $k(1 \leq k \leq c)$. If each released statistics for $k$th-hop($1 \leq k \leq c$) connection fingerprints preserves $\mathcal{P}/c$ (or $\mathcal{P}/2^k$)-PDP, the series of $c$ counting queries guarantee $\mathcal{P}$-PDP by theorem 1. These two baseline methods are denoted as \emph{Uniform} and \emph{Exponential}, correspondingly. Both \emph{Uniform} and \emph{Exponential} methods are easy to achieve $\mathcal{P}$-PDP, but they ignore the fact that the statistics may not change significantly for successive queries $f_k$ due to the sparsity of social network and inherits a large quantity of unnecessary noises.

\textbf{Distance-based method.} In this paper, we use a distance-based budget allocation approach inspired by \cite{30} to reduce noise. Our proposed DEBA starts by distributing the publication budget in an exponentially decreasing fashion to every private counting query $f_k(1 \leq k \leq c)$, i.e., query $f_k$ receives $P/2^{k+1}$ to publish its counting results. If it is found out that the distance between the statistics for $f_k$ and $f_{k-1}(2 \leq k \leq c-1)$ is smaller than its publication threshold, the counting query $f_k$ is skipped and its corresponding publication budget becomes available for a future counting query. On the other hand, if it is decided to publish the counting results of $f_k$ , the $f_k$ should absorb all the budgets that became available from the previous skipped counting queries, and uses it in order to publish the current counting query $f_k$ with higher accuracy.

The presence or absence of one edge in the graph can contribute to a large number of potential CFPs, i.e., the global sensitivity of the counting queries is large and so the noise added to the count has to be scaled up. Our second method, DUBA-LF, further improves DEBA and uses a new technique, called ¡°ladder functions¡±, for producing differentially private output. The technique specifies a probability distribution over possible outputs that are carefully defined to maximize the utility for the given input, while still providing the required privacy level. Moreover, we start the DUBA-LF by uniformly distributing the budget instead of the exponentially distributing in DEBA.

\section{Proposed Methods}
We propose a distance-based budget absorption approach to release the number of $k$th-hop($1 \leq k \leq c$) CFPs under $\mathcal{P}$-PDP. Instead of releasing a $\mathcal{P}/c$ (or $\mathcal{P}/2^k$)-PDP result for every $k(1 \leq k \leq c)$, the new publication results are computed if and only if the distance between the counting statistics and the latest released statistics is larger than a threshold. It is worth noting that the statistics may not change significantly for successive queries $f_k$ due to the sparsity of social network. Therefore, this distance-based budget allocation approach can save some privacy budgets for a future counting query and reduce the overall error of released statistics.

In this section, our basic method, called DEBA, is presented at first. The basic method starts by \emph{exponentially} distributing the budget to every private counting query $f_k (1 \leq k \leq c)$, and the budget absorption is decided by the distance between the counting statistics and the latest released statistics.
We then introduce our advanced method, DUBA-LF, which uses ladder function to reduce the noise introduced by the traditional differentially private mechanism.

\subsection{DEBA}
DEBA(Publication with \underline{D}istance-based \underline{E}xponential \underline{B}udget \underline{A}bsorption) starts with an exponentially decreasing budget for every private counting query $f_k (1 \leq k \leq c)$, and then a privacy-preserving distance calculation mechanism is adopted to measure the distance between the counting statistics and the latest released statistics. The decision step uses the distance to decide whether to publish the private counting results of $f_k$ or not. If the decision is not, the private counting results of $f_k$ are approximated with the last \emph{non-null} publication and the budget of $f_k$ becomes available for a future counting query. Otherwise, the private counting query $f_k$ absorbs all the budgets that are available from the previous skipped counting queries. The overall privacy budget is divided between the decision and publishing steps which are designed to guarantee personalized differential privacy as we will analyze later.

Before introducing the proposed DEBA, we give the sensitivity of counting query $f_k (1 \leq k \leq c)$ at first. Neighbor graphs of $G$ are all the graphs $G'$ which differ from $G$ by at most a single edge. For the counting query $f_1$, it queries the number of 1st-hop connection fingerprints for each private user in the sensitive graph, and changing a single edge in $G$ will result in at most one entry changing in the 1st-hop connection fingerprints. Hence, $\Delta(f_1)=1$. For the counting query $f_k (2 \leq k \leq c)$, changing a single edge in $G$ will result in at most $|V_{pub}| = m_p$ entries changing in the $k$th-hop connection fingerprints, i.e., $\Delta(f_k)= m_p$ for $2 \leq k \leq c$.

\begin{algorithm}[htbp]
\caption{Pseudocode of DEBA}
\label{alg:deba}
\setlength{\abovecaptionskip}{0.cm}
\setlength{\belowcaptionskip}{-0.cm}
\begin{algorithmic}[1]
\REQUIRE ~~\\
$G -$ private input graph\\
$V_{pri,G} -$ the set of private users in $G$\\
$V_{pub,G} -$ the set of public users in $G$\\
$\mathcal{P} -$ the privacy specification\\
$m -$ the cardinality of $V_{pri,G}$\\
$c -$ the counting range\\
$t -$ the configurable sample threshold\\
\ENSURE ~~\ $\tilde F = {\rm{(}}{\tilde F_1}{\rm{,}}{\tilde F_2}{\rm{,}}......{\rm{,}}{\tilde F_c}{\rm{)}}$;

\textbf{For} each $k$($1 \leq k \leq c$) \textbf{do}\\
//\emph{personal private distance calculation mechanism $M_1$}
\begin{enumerate}[1{:}]
\STATE Calculate $F_k= f_k(G)$.
\STATE Identify last \emph{non-null} release ${\tilde F_r}$ from ${\tilde F}$.
\STATE Sample $c_k=RS(F_k ,\mathcal{P}/2c, t/2c)$ with probability $\pi(v_i,t/2c)$.
\STATE Set $dist = \frac{1}{m}\sum\nolimits_{j = 1}^m {\left| {{{\tilde F}_r}{\rm{[}}j{\rm{]}} - {c_k}{\rm{[}}j{\rm{]}}} \right|} $ and $\epsilon_{k,1}= t/2c$.
\STATE Calculate $dist= dist +Lap(2m_pc/mt)$.\\
//\emph{personal private publication mechanism $M_2$}
\STATE \textbf{if} $k=1$ \textbf{then}
\STATE Set $\epsilon_{k,2}= t/4$.
\STATE Sample $G_k=RS(G,\mathcal{P}/4,t/4)$ with probability $\pi(e_{ij},t/4)$.
\STATE Calculate  ${\tilde F_k} = {\tilde f_k}{\rm{(}}G{\rm{) = }}{f_k}{\rm{(}}{G_k}{\rm{) + Lap(}}\frac{4}{t}{\rm{)}}$.
\STATE \textbf{else if} $k<c$ \textbf{then}
\STATE Calculate $\epsilon_{k,2}= \sum\nolimits_{j = r + 1}^k {\frac{t}{{{2^{j + 1}}}}} $ and $T_k = m_p/\epsilon_{k,2}$.
\STATE \textbf{if} $dist > T_k$, \textbf{then}
\STATE Sample $G_k=RS(G,\mathcal{P} \cdot \sum\nolimits_{j = r + 1}^k {\frac{1}{{{2^{j + 1}}}}},\epsilon_{k,2})$ with probability $\pi(e_{ij},\epsilon_{k,2})$.
\STATE Calculate  ${\tilde F_k} = {\tilde f_k}{\rm{(}}G{\rm{) = }}{f_k}{\rm{(}}{G_k}{\rm{) + Lap(}}\frac{m_p}{\epsilon_{k,2}}{\rm{)}}$.
\STATE \textbf{end if}
\STATE \textbf{else if} $k=c$ \textbf{then}
\STATE Calculate $\epsilon_{k,2}= \sum\nolimits_{j = r + 1}^k {\frac{t}{{{2^{j + 1}}}}} $ and $T_k = m_p/\epsilon_{k,2}$.
\STATE Sample $G_k=RS(G,\mathcal{P} \cdot \sum\nolimits_{j = r + 1}^k {\frac{1}{{{2^{j + 1}}}}},t/4)$ with probability $\pi(e_{ij},\epsilon_{k,2})$.
\STATE Calculate  ${\tilde F_k} = {\tilde f_k}{\rm{(}}G{\rm{) = }}{f_k}{\rm{(}}{G_k}{\rm{) + Lap(}}\frac{m_p}{\epsilon_{k,2}}{\rm{)}}$.
\STATE \textbf{else} set $\tilde F_k= null$.
\STATE \textbf{end if}\\
\textbf{EndFor}
\end{enumerate}
\end{algorithmic}
\end{algorithm}

Algorithm 1 presents the pseudocode of DEBA. DEBA is decomposed into two sub mechanisms: \emph{personal private distance calculation mechanism $M_1$} and \emph{personal private publication mechanism $M_2$}. Line 1-5 capture the calculation of personal private distance between the counting statistics and the latest released statistics, labeled as mechanism $M_1$. Line 6-9 carry out the publication step for 1st-hop connection fingerprints and line 10-21 carry out the publication step for $k$th-hop$(2 \leq k \leq c)$ connection fingerprints. Line 11 or line 17 gets the total budgets of skipped queries whose budgets for publication is absorbed. Then the publication threshold $T_k$ for query $f_k$ is determined by $m_p/\epsilon_{k,2}$. The reason to define such a threshold is that the injecting Laplace noise of $f_k$ is with scale $T_k$. Then DEBA compares the distance $dist$ to the threshold $T_k$(line 12). If the distance is larger than $T_k$ , DEBA samples the private social network $G$(Line 13) and outputs the noisy counts (Line 14), or $null$ otherwise (Line 20). In addition, DEBA outputs $c$th-hop connection fingerprints with the totally remaining budgets as shown in Line 16-19.

\textbf{Remark:} Recall that the error of randomly sampling input graph $G$ is data-dependent as well as the error of distance based approximation. And we cannot present a formal utility analysis for such a data-dependent mechanism. We will present extensive experiments using real datasets to justify the performance of our algorithms. Moreover, precisely optimizing $t$ for an arbitrary $f$ may be nontrivial in practice because, although the error of $DP_{f}^{t}$ may be quantified without knowledge of the dataset, the impact of sampling does depend on the input data. A possible option, in some cases, is to make use of old data that is no longer sensitive (or not as sensitive), and that comes from a similar distribution, to approximately optimize the threshold without violating privacy. It is demonstrated that for many functions, the simple heuristics of setting $t = \max_v P^v$ or $t = \frac{1}{m} \sum_{v} P^v$ , often give good results on real data and privacy specifications\cite{6}.

\subsection{DUBA-LF}

The proposed DEBA mechanism publishes the private count of $f_k$ by adding Laplace noise to the true answer, where the scale of noise is proportional to the global sensitivity of $f_k$. It is pointed out that the global sensitivity of $f_k$ is 1 for $k =1$ or $m_p$ for $2 \leq k \leq c$. It is obvious that there can be numerous public users in large network graphs. Hence, the global sensitivity of counting query $f_k$ may be very large and makes the noise large enough to overwhelm the true answer. In order to improve the utility of private release for $f_k$, we use the new definition of ladder function to reduce the introduced noise. The definition of ladder function is presented at first.
\begin{definition}
(Ladder function\cite{25}). A function $I_x(G)$ is said to be a ladder function of query $f$ if and only if
\\$  (a) LS (G, f) \leq I_0(G)$, for any $G$;
\\$  (b) I_x(G') \leq I_{x+1}(G)$, for any pair of neighboring graphs $G, G'$, and any nonnegative integer $x$.
\end{definition}

A straightforward example of a ladder function for count query $f_k$ is $I_t(G, f_k) = \Delta(f_k)$, since $LS (G, f) \leq \Delta(f_k)$ for any $G$, and a constant always satisfies the second requirement. However, as aforementioned, the global sensitivity of counting query $f_k$ can be extremely large for CFP counting, which may not require so much noise.

For counting query $f_1$, its global sensitivity is 1 and the ladder function for $f_1$ can be defined as $I_t(G, f_1) = 1$. Before detailing the ladder function for $f_k (2 \leq k \leq c)$, the important notation local sensitivity is refined by defining the sensitivity for a particular pair of nodes by defining the sensitivity for a particular pair of nodes $(v_i,v_j)$, denoted by $LS_{ij}(G,f)$. Then $LS(G, f) = \max_{i,j} LS_{ij}(G, f)$. Let $p_i$ denotes the number of the number of 1st-hop connection fingerprints for user $v_i$ and $d_{max}$ be the maximum node degree in $G$. Then it is easy to get that $LS(G, f_k) = \max_i p_i$ for $f_k (2 \leq k \leq c)$.Without losing of generality, we simply assume that $p_i \leq p_j$ for a particular pair of nodes $(v_i,v_j)$. Then we give our ladder function for $k$th-hop connection fingerprint counting queries $f_k (2 \leq k \leq c)$ in Theorem 4 and prove that the constructed ladder function satisfy the requirements in Definition 10.

\begin{theorem}
$I_t(G, f_k) = \min\{m_p,LS(G, f_k)+t\} $is a ladder function for $f_k (2 \leq k \leq c)$.
\end{theorem}
\begin{IEEEproof}
The proof contains the following two steps.
\\    $(i)LS(G, f_k ) \leq I_0(G, f_k)$ for any $G$. This step is trivial since $I_0(G, f_k) = LS(G, f_k)$.
\\	$(ii)I_t(G', f_k) \leq I_{t+1}(G, f_k)$ for any neighboring graphs $G'$ and $G$. Note the fact that the set $\{G^* |d(G^*,G')\leq t\}$ is a subset of $\{G^* |d(G^*,G)\leq t+1\}$. Therefore, $\max_{G^*|d(G^*,G')\leq t}\| f_k(G')-f_k (G^*)\| \leq \max_{G^*|d(G^*,G)\leq t+1} \|f_k (G^* )-f_k(G )\|$, i.e., $I_{t+1}(G, f_k)
= \min\{m_{p},LS(G, f_k)+t+1\} = \min\{m_{p},LS(G, f_k,t+1)\} \geq = \min\{m_{p},LS(G', f_k,t)\} = I_t(G', f_k)$.
\end{IEEEproof}

It is clear that the ladder function $I_t(G, f_k)$ converges to $\Delta(f_k)$ when $t\geq \Delta(f_k) - LS(G, f_k)$. The ladder function $I_t(G, f_k)$ is used to determine the quality function $q$ in exponential mechanism and define how $q$ varies. In particular, $q$ is a symmetric function over the entire integer domain, centered at $f_k(G)$. The quality function $q$ is defined as follows:
\begin{definition}
(Ladder Quality\cite{25}). Formally, given ladder function $I_x(G, f_k)$ we define the ladder quality function $q_{f_k}(G,v_i,s)$ for node $v_i$ by
\\ $ (i) \ \ q_{f_k}(G, v_i , f_k(v_i)) = 0$;
\\ $ (ii)$ for $s \in f_k(v_i)\pm(\sum_{t=0}^{u-1} I_x(G, f_k), \sum_{t=0}^{u} I_t(G, f_k)]$, set $q_{f_k}(G, v_i ,s) = -u-1$.
\\After assigning each integer a quality score, the sensitivity of the quality function can be calculated as, $\Delta(q_{f_k}) = \max_{v_i,G,G'} \|q_{f_k}(G,v_i,s)-q_{f_k}(G',v_i,s)\| =1$. We refer the reader to \cite{25}(THEOREM 4.2) for a full description of the proof of it.
\end{definition}

\begin{algorithm}[htbp]
\caption{Pseudocode of DUBA-LF}
\label{alg:DUBA-LF}
\setlength{\abovecaptionskip}{0.cm}
\setlength{\belowcaptionskip}{-0.cm}
\begin{algorithmic}[1]
\REQUIRE ~~\\
$G -$ private input graph\\
$V_{pri,G} -$ the set of private users in $G$\\
$V_{pub,G} -$ the set of public users in $G$\\
$\mathcal{P} -$ the privacy specification\\
$m -$ the cardinality of $V_{pri,G}$\\
$c -$ the counting range\\
$t -$ the configurable sample threshold\\
\ENSURE ~~\ $\tilde F = {\rm{(}}{\tilde F_1}{\rm{,}}{\tilde F_2}{\rm{,}}......{\rm{,}}{\tilde F_c}{\rm{)}}$;

\textbf{For} each $k$($1 \leq k \leq c$) \textbf{do}\\
//\emph{personal private distance calculation mechanism $M_1$}\\
//\emph{Same as Lines 1-5 in Algorithm 1}\\
//\emph{personal private publication mechanism $M_2$}\\
\begin{enumerate}[1{:}]
\setcounter{enumi}{5}
\STATE \textbf{if} $k=1$ \textbf{then}
\STATE Set $\epsilon_{k,2}= t/2c$.
\STATE Sample $G_k=RS(G,\mathcal{P}/2c,t/2c)$ with probability $\pi(e_{ij},t/2c)$.
\STATE Calculate  ${\tilde F_k} = {\tilde f_k}{\rm{(}}G{\rm{) = }}{f_k}{\rm{(}}{G_k}{\rm{) + Lap(}}\frac{2c}{t}{\rm{)}}$.
\STATE \textbf{else if} $k<c$ \textbf{then}
\STATE Calculate $\epsilon_{k,2}= (k-r)t/2c$ and $T_k = m_p/\epsilon_{k,2}$.
\STATE \textbf{if} $dist > T_k$, \textbf{then}
\STATE Sample $G_k=RS(G,\mathcal{P} \cdot (k-r)/c,\epsilon_{k,2})$ with probability $\pi(e_{ij},\epsilon_{k,2})$.
\STATE Calculate  ${\tilde F_k} = LFNoising(f_k(G_k),\epsilon_{k,2}, I_x(G,f_k))$.
\STATE \textbf{end if}
\STATE \textbf{else if} $k=c$ \textbf{then}
\STATE Calculate $\epsilon_{k,2}= (k-r)t/2c$ and $T_k = m_p/\epsilon_{k,2}$.
\STATE Sample $G_k=RS(G,\mathcal{P} \cdot (k-r)/c,\epsilon_{k,2})$ with probability $\pi(e_{ij},\epsilon_{k,2})$.
\STATE Calculate  ${\tilde F_k} = LFNoising(f_k(G_k),\epsilon_{k,2}, I_x(G,f_k))$.
\STATE \textbf{else} set $\tilde F_k= null$.
\STATE \textbf{end if}\\
\textbf{EndFor}
\end{enumerate}
\end{algorithmic}
\end{algorithm}

\begin{algorithm}[htbp]
\caption{\emph{LFNoising}$(f_k(G_k),\epsilon_{k,2}, I_x(G,f_k))$}
\label{alg:LFNoising}
\setlength{\abovecaptionskip}{0.cm}
\setlength{\belowcaptionskip}{-0.cm}
\begin{algorithmic}[1]
\REQUIRE ~~\\
$f_k(G_k) -$ the number of $k$th-hop connection fingerprints in sampled input graph $G_k$\\
$\epsilon_{k,2} -$ the notional publication budget\\
$I_x(G,f_k) -$ the ladder function\\

\ENSURE ~~\ ${\tilde F_k} = {\rm{(}}{\tilde F_k}{\rm{[1],}}{\tilde F_k}{\rm{[2],}}......{\rm{,}}{\tilde F_k}{\rm{[}}m{\rm{])}}$;

\begin{enumerate}[1{:}]
\STATE Set $d=0$, $range[0] = f_k(G_k)$ and $weight[0] = {\rm{exp(}}\frac{{{\varepsilon _{k,2}}}}{2} \cdot {\rm{0)}}$.
\STATE \textbf{for} $x=1$ to $M = m_p-LS(G,f_k)$ \textbf{do}
\STATE $range[x] = f_k(G_k)\pm(d, d+ I_{x-1}(G, f_k)]^m$.
\STATE $weight[x] = 2I_{x-1}(G, f_k){\rm{exp(}}\frac{{{\varepsilon _{k,2}}}}{2} \cdot {\rm{(}} - x{\rm{))}}$.
\STATE $d = d + I_{x-1}(G, f_k)$.
\STATE \textbf{end for}
\STATE $weight[M+1] = \frac{{{\rm{2}}{m_p}{\rm{exp(}}\frac{{{\varepsilon _{k,2}}}}{2} \cdot {\rm{(}} - M - 1{\rm{))}}}}{{1 - {\rm{exp(}} - \frac{{{\varepsilon _{k,2}}}}{2}{\rm{)}}}}$.
\STATE Randomly sample $m$ integers $T = (t_1, t_2,\cdots, t_m)$. Here $t_i (1 \leq i \leq m)$is draw with probability $weight[t_i]$ over sum of weights.
\STATE \textbf{for} {$i =1$ to $m$} \textbf{do}
\STATE \textbf{if} $t_i \leq M$ \textbf{then} uniformly sample an integer $j$ from $range[t_i][i]$ and set ${\tilde F_k}{\rm{[}}i{\rm{]}}=j$;
\STATE \textbf{else} sample an integer $h$ from the geometric distribution with parameter $p = 1- {\rm{exp(}} - \frac{{{\varepsilon _{k,2}}}}{2}{\rm{)}}$.
\STATE uniformly sample an integer $j$ from $f_k(G_k)[i]\pm{d+hm_p+(0,m_p]}$ and set ${\tilde F_k}{\rm{[}}i{\rm{]}}=j$.
\STATE \textbf{end if}
\STATE \textbf{end for}
\RETURN ${\tilde F_k}$.
\end{enumerate}
\end{algorithmic}
\end{algorithm}

DUBA-LF (Publication with \underline{D}istance-based \underline{U}niformly \underline{B}udget \underline{A}bsorption using \underline{L}adder \underline{F}unction) uses ladder function to reduce the introduced noise while reallocating the pre-allocated uniform privacy budget. The pseudocode of DUBA-LF is presented in Algorithm 2. \emph{The personal private distance calculation mechanism $M_1$} is identical to that of DEBA (Lines 1-5 in Algorithm 1). \emph{The personal private publication mechanism $M_2$} is presented in Lines 6-21. Lines 6-9 carry out the publication step for counting query $f_1$. The publication step for $k$th-hop($2 \leq k \leq c$) connection fingerprints is carried out in Line 10-21. DUBA-LF samples the private social network $G$(Line 13 or Line 18) in the same way with DEBA but the sampling probabilities are different. The personal private publication for $k$th-hop($2 \leq k \leq c$) connection fingerprints in DUBA-LF is also different to DEBA. If the distance is larger than $T_k$ or counting for $f_c$, DUBA-LF uses an exponential mechanism based mechanism \emph{LFNoising} to provide differential privacy (Line 14 and Line 19). In the meantime, DUBA-LF outputs $null$ if the distance is not larger than $T_k$ (Line 20).

\emph{LFNoising} is an extending algorithm of \emph{NoiseSample} in \cite{25}. \emph{NoiseSample} is proposed to output one value as the final differentially private result while our proposed \emph{LFNoising} is aims to solve the problem of differentially private releasing in the vector form. The pseudocode of \emph{LFNoising} is presented in Algorithm 3. Given the ladder function $I_t(G,f_k)$, the calculation of the range and weight for the first few rungs, e.g., rung 0 (the center) to rung $M+1$ ($M = m_p - LS(G,f_k)$)are shown in Lines 1-7. Lines 8-12 describe the random sampling of the private publication vector ${\tilde F_k}$ which presents the private count of $k$th-hop($2 \leq k \leq c$) connection fingerprints.

\section{Privacy Analysis}
The proofs of privacy guarantees for the proposed mechanisms are formally provided in this section. We show the proposed DEBA satisfies P-personalized differential privacy first.

\begin{lemma}
Mechanism $M_1$ in Algorithm 1 is $\mathcal{P}/2$-personalized differentially private.
\end{lemma}
\begin{IEEEproof}
We use the notation $F_{k,-v}$ and $F_{k,+v}$ to mean the graph resulting from removing from or adding to $F_k$ the tuple $f_k(v)$. We can represent two neighboring datasets (vectors) as $F_k$ and $F_{k,-v}$. For each $1 \leq k \leq c$, all of the possible outputs of $RS(F_k,\mathcal{P}/2c, t/2c)$ can be divided into those in which $f_k(v)$ was selected, and those in which $f_k(v)$ was not selected. The sensitivity of $dist$ function in $M_1$ is $m_p/m$, therefore, the mechanism injects Laplace noise with scale $2m_pc/mt$ in Line 5 can be denoted as $DP_{dist}^{t/2c}$.Thus, we have\\
\begin{equation}
\begin{aligned}
\small
&Pr[{S_{{f_k}}}{\rm{(}}{F_k},\mathcal{P}/2c,t/2c{\rm{)}} \in O]\\
&= Pr[DP_{dist}^{t/2c}(RS(F_k,\mathcal{P}/2c,t/2c))\in O] \\
&= \sum_{Z \in F_{k, -v}} (\pi(v,t/2c)Pr[RS(F_k,\mathcal{P}/2c,t/2c)= Z]\cdot \\
&Pr[DP_{dist}^{t/2c}(Z_{+ v})\in O]) + \sum_{Z \in F_{k, - v}} ((1 - \pi(v,t/2c))\cdot \\
&Pr[RS(F_k,\mathcal{P}/2c,t/2c)= Z]Pr[DP_{dist}^{t/2c}(Z) \in O]) \\
&\leq \sum_{Z \in F_{k, - v}}(\pi(v,t/2c)Pr[RS(F_k,\mathcal{P}/2c,t/2c) = Z]e^{t/2c}\cdot \\
&Pr[DP_{dist}^{t/2c}(Z) \in O])+ \\
&(1 - \pi(v,t/2c))S_{f_k}(F_{k, - v},\mathcal{P}/2c,t/2c) \\
&\leq e^{t/2c}\pi(v,t/2c)S_{f_k}(F_{k,-v},\mathcal{P}/2c,t/2c)+ \\
&(1-\pi(v,t/2c))S_{f_k}(F_{k,-v},\mathcal{P}/2c,t/2c) \\
&= (1 - \pi (v,t/2c) + e^{t/2c}\pi(v,t/2c))S_{f_k}(F_{k, - v},\mathcal{P}/2c,t/2c)
\end{aligned}
\end{equation}
There are two cases for $v$ that we must consider: (1)$P^v/2c \geq t/2c$; (2)$P^v/2c < t/2c$. For the former case, we have $\pi(v,t/2c)=1$ and eq.(1) can be rewritten as\\
\begin{equation*}
\begin{aligned}
\small
&Pr[{S_{{f_k}}}{\rm{(}}{F_k},\mathcal{P}/2c,t/2c{\rm{)}} \in O]\\
&\leq (1 - 1 + e^{t/2c}\cdot 1)S_{f_k}(F_{k, - v},\mathcal{P}/2c,t/2c)\\
&= e^{t/2c}S_{f_k}(F_{k, - v},\mathcal{P}/2c,t/2c)\\
&\leq e^{P^v/2c}S_{f_k}(F_{k, - v},\mathcal{P}/2c,t/2c)
\end{aligned}
\end{equation*}
For the latter case $P^v/2c < t/2c$,\\
\begin{equation*}
\begin{aligned}
\small
&Pr[{S_{{f_k}}}{\rm{(}}{F_k},\mathcal{P}/2c,t/2c{\rm{)}} \in O]\\
&\leq (1 - \pi (v,t/2c) + e^{t/2c}\pi(v,t/2c))S_{f_k}(F_{k, - v},\mathcal{P}/2c,t/2c)\\
&= {\rm{ (1}} - \frac{{{e^{{P^v}/2c}} - 1}}{{{e^{t/2c}} - 1}}{\rm{ + }}{e^{t/2c}}\frac{{{e^{{P^v}/2c}} - 1}}{{{e^{t/2c}} - 1}}{\rm{)}}{S_{{f_k}}}{\rm{(}}{F_{k, - v}},P/2c,t/2c{\rm{)}}\\
&= \frac{{{e^{{P^v}/2c}}{\rm{(}}{e^{t/2c}} - 1{\rm{)}}}}{{{e^{t/2c}} - 1}}{S_{{f_k}}}{\rm{(}}{F_{k, - v}},P/2c,t/2c{\rm{)}}\\
&= e^{P^v/2c}S_{f_k}(F_{k, - v},\mathcal{P}/2c,t/2c)
\end{aligned}
\end{equation*}
To sum up, we have $Pr[{S_{{f_k}}}{\rm{(}}{F_k},\mathcal{P}/2c,t/2c{\rm{)}} \in O] \leq \\
e^{P^v/2c}S_{f_k}(F_{k, - v},\mathcal{P}/2c,t/2c)$, and for each $1 \leq k \leq c$, the mechanism satisfies $\mathcal{P}/2c$-PDP. Therefore, according to Theorem 2, mechanism $M_1$ in Algorithm 1 is $\mathcal{P}/2$-personalized differentially private.
\end{IEEEproof}

We have proved that mechanism $M_1$ satisfies $\mathcal{P}/2$-personalized differential privacy. To prove that DEBA satisfies $\mathcal{P}$-personalized differential privacy, we must prove that, for every $k (1 \leq k \leq c)$, $M_2$ is $\mathcal{P}\cdot\sum\nolimits_{j = r + 1}^k {\frac{t}{{{2^{j + 1}}}}} $ - personalized differentially private if it publishes, and 0 - personalized differentially private otherwise.

\begin{theorem}
The proposed DEBA satisfies $\mathcal{P}$-PDP.
\end{theorem}
\begin{IEEEproof}
 Mechanism $M_1$ satisfying $\mathcal{P}/2$-personalized differential privacy is captured in Lemma 1. Mechanism $M_2$ publishes ${\tilde F_k}$ or $null$. In the latter case, the privacy budget is trivially equal to zero, as no publication occurs. In the former case, the sensitivity of $f_k$ is mp for $2 \leq k \leq c$ and 1 for $k =1$ and the publication budget depends on previous publications. Hence, the mechanism injects Laplace noise with scale $\frac{{{m_p}}}{{\sum\nolimits_{j = r + 1}^k {\frac{t}{{{2^{j + 1}}}}} }}$ can be denoted as $DP_{{f_k}}^{\sum\nolimits_{j = r + 1}^k {\frac{t}{{{2^{j + 1}}}}} }$ for $2 \leq k \leq c$ and the mechanism injects Laplace noise with scale $4/t$ can be denoted as $DP_{{f_1}}^{t/4}$ for $k =1$.Following the proof technology in Lemma 1, it is easy to prove that $M_2$ is $\mathcal{P}\cdot\sum\nolimits_{j = r + 1}^k {\frac{t}{{{2^{j + 1}}}}} $-PDP if it publishes a \emph{non-null} result for each $k (1 \leq k \leq c)$. Moreover, the total publication budget is $\mathcal{P}/2$, and it at most equals to the case where each of these $c$ queries receives a budget of $\frac{t}{{{2^{k + 1}}}}$. So, $\sum\limits_{j = 1}^c {\frac{1}{{{2^{j + 1}}}}} \mathcal{P} \le \frac{\mathcal{P}}{2}$. According to Theorem 2, we get the conclusion that $M_2$ satisfies $\mathcal{P}/2$-PDP. To sum up, the proposed DEBA satisfies $\mathcal{P}$-PDP.
\end{IEEEproof}

DUBA-LF employs a personal private distance calculation mechanism $M_1$ identical to that of DEBA and its privacy guarantee is captured by Lemma 1. In order to show the mechanism $M_2$ in DUBA-LF satisfies $\mathcal{P}/2$-PDP, we need to prove that the algorithm \emph{LFNoising}$(f_k(G_k),\epsilon_{k,2}, I_t(G,f_k))$ is $\epsilon_{k,2}$- differentially private, i.e., \emph{LFNoising}$(f_k(G_k),\epsilon_{k,2}, I_t(G,f_k))$ can be denoted as $DP_{{f_k}}^{{\varepsilon _{k,2}}}$.
\begin{lemma}
\emph{LFNoising}$(f_k(G_k),\epsilon_{k,2}, I_t(G,f_k))$ is $\epsilon_{k,2}$- differentially private.
\end{lemma}
\begin{IEEEproof}
There are two steps in the algorithm \emph{LFNoising}: selecting a rung of the ladder (where rung $M+1$ is considered as a special case) according to the relative value of the weight of the rung and picking an integer from the corresponding rung. For rungs 0 to $M$, the possible output values on the same rungs are picked uniformly. For rung $M+1$, the possible outputs are determined by two actions£º picking how many further rungs down the ladder to go and then picking uniformly from these. As discussed above, for $1 \leq i \leq m$, the output probability distribution is equal to\\
$Pr[{\tilde F_k}{\rm{[}}i{\rm{] = }}\rho] = \frac{{\exp {\rm{(}}\frac{{{\varepsilon _{k,2}}}}{{2\Delta \left( {{q_{{f_k}}}} \right)}} \cdot {q_{{f_k}}}{\rm{(}}G,{v_i}{\rm{,}}\rho {\rm{))}}}}{{\sum\limits_{\rho  \in } {\exp {\rm{(}}\frac{{{\varepsilon _{k,2}}}}{{2\Delta \left( {{q_{{f_k}}}} \right)}} \cdot {q_{{f_k}}}{\rm{(}}G,{v_i}{\rm{,}}\rho {\rm{))}}} }}$\\
As argued above, if the input graph $G$ is replaced by its neighboring graph $G¡ä$, the quality of $\rho$ will be changed by at most $\Delta \left( {{q_{{f_k}}}} \right) = 1$, i.e., the numerator $\exp {\rm{(}}\frac{{{\varepsilon _{k,2}}}}{{2\Delta \left( {{q_{{f_k}}}} \right)}} \cdot {q_{{f_k}}}{\rm{(}}G,{v_i}{\rm{,}}\rho {\rm{))}}$ can change at most $\exp {\rm{(}}\frac{{{\varepsilon _{k,2}}}}{{2\Delta \left( {{q_{{f_k}}}} \right)}} \cdot \Delta \left( {{q_{{f_k}}}} \right){\rm{) = }}\exp {\rm{(}}\frac{{{\varepsilon _{k,2}}}}{2}{\rm{)}}$. Moreover, a single change in graph $G$ the changing in denominator is minimized by a factor of $\exp {\rm{(}} - \frac{{{\varepsilon _{k,2}}}}{2}{\rm{)}}$, giving the ratio of the new probability of $\rho$ and the original one $\exp {\rm{(}}{\varepsilon _{k,2}}{\rm{)}}$. Therefore,\emph{LFNoising}$(f_k(G_k),\epsilon_{k,2}, I_t(G,f_k))$ is $\epsilon_{k,2}$- differentially private.
\end{IEEEproof}

This result highlights the fact that \emph{LFNoising}$(f_k(G_k),\epsilon_{k,2}, I_t(G,f_k))$ can be denoted as $DP_{{f_k}}^{{\varepsilon _{k,2}}}$. Similar to Theorem 5, we can conclude that DUBA-LF is $\mathcal{P}$-personalized differentially private.
\begin{theorem}
The proposed DUBA-LF satisfies $\mathcal{P}$-PDP.
\end{theorem}
\begin{IEEEproof}
The proof is similar to that of Theorem 5 and we omit it.
\end{IEEEproof}

\section{Experimental Evaluation}

We make use of three real-world graph datasets in our experiments: \textbf{polblogs}\cite{31}, \textbf{facebook}\cite{32}
and \textbf{CondMat}\cite{33} networks. The \textbf{polblogs} network was crawled from the US political blogosphere in 2005. The vertices are blogs of a set of US politicians, and an edge between two blogs represents the existence of hyperlinks from one blog to the other. The \textbf{facebook} network was collected from the survey participants using a Facebook app. The vertices are Facebook users, and an edge between two users represents the established friendship between them. The \textbf{CondMat} network was collaboration networks from the e-print arXiv, which cover scientific collaborations between authors who submitted papers to Condensed Matter category. The edge between two authors represents an author co-authored a paper with another author in this network. All the networks are represented by undirected and unweighted graphs with no isolated vertices.

The real-world networks we used do not contain public user identity. In other words, all vertices in the networks are anonymous. In order to evaluate the proposed CFPs publication algorithms, we select a set of vertices in each network assuming their identities are public. Thereafter, based on these public vertices, we generate the CFPs of the remaining private vertices. In this paper, we select vertices with the highest degrees as public vertices and the proportion of public users is set to 5\% in our whole experiment. Table I presents some basic statistics of the networks.
\begin{table}[!t]
\renewcommand{\arraystretch}{1.3}
\caption{Datasets Properties}
\label{table_example}
\centering
\begin{tabular}{c|c|c|c|c|c}
\hline
\bfseries Dataset & \bfseries nodes & \bfseries edges &  \bfseries diameter & \bfseries density & $\bm{|V_{pub}|}$\\
\hline\hline

Polblogs & 1222	& 16724	& 8	& 2.24\% &	61\\
Facebook & 4039	& 88234	& 8	& 1.08\%	& 201\\
CondMat & 23133 & 93439 & 14 & 0.0349\% & 1156\\
\hline
\end{tabular}
\end{table}

We compared DEBA and DUBA-LF with benchmarks \emph{Uniform} and \emph{Exponential} over these three datasets. We implemented all methods in Matlab, ran each experiment 100 times, and reported the average error, expressed as the Mean Absolute Error (MAE) and the Mean Relative Error (MRE). To generate the privacy specifications for our experiments, we randomly divided the private users (records) into three groups: conservative, representing users with high privacy concern; moderate, representing users with medium concern; and liberal, representing users with low concern. The fraction of each type users is 1/3. The privacy preferences for the users in the conservative, moderate and liberal groups received a privacy preference of $\epsilon_c = 1$, $\epsilon_m = 4$ and $\epsilon_l = 16$ respectively. As a result, the average privacy preference of all users equals to 7.

\begin{figure}[h]
  \setlength{\abovecaptionskip}{0.cm}
  \setlength{\belowcaptionskip}{-0.5cm}
  \centering
  \includegraphics[scale = 0.46]{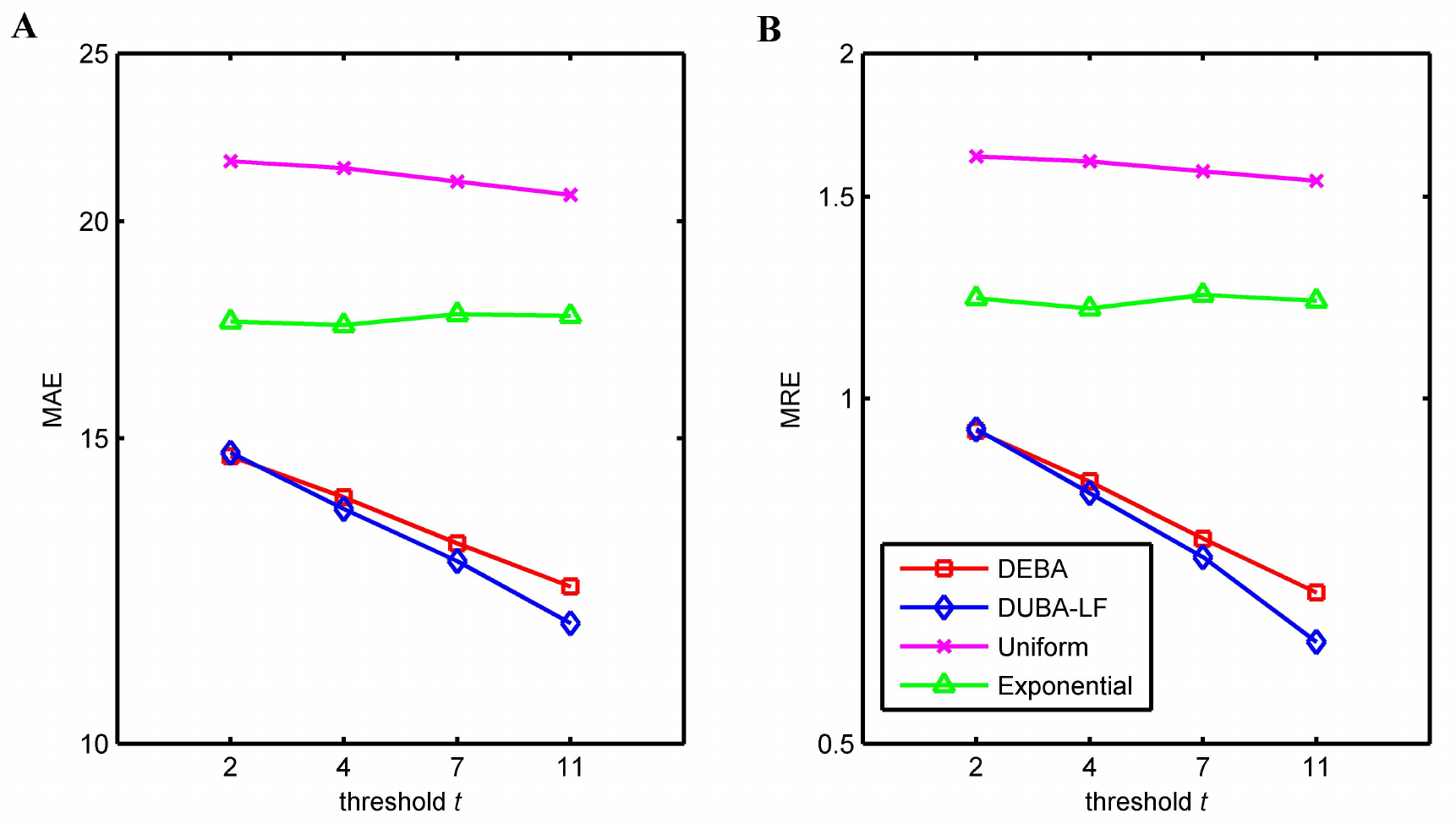}
  \caption{MAE and MRE of each mechanism for the Polblogs dataset, as sample threshold is varied and $c$ is set to 4.}
\end{figure}

Fig.1 plots the MAE and MRE of all schemes for the Polblogs dataset, where we vary the sampling threshold $t$ and set $c$ = 4. DUBA-LF is the best method in this setting. And it outperforms \emph{Uniform} mechanism by up to 76.6\% in MAE and 152.5\% in MRE, \emph{Exponential} by up to 50.3\% in in MAE and 98.4\% in MRE, and DEBA by up to 5.0\% in MAE and 10.4\% in MRE. The results also indicate that increasing the sampling threshold has an effect of decreasing of MAR and MRE for DUBA-LF, DEBA and \emph{Uniform} mechanisms but the effect for \emph{Exponential} mechanism is not evident.

\begin{figure}[h]
  \setlength{\abovecaptionskip}{0.cm}
  \setlength{\belowcaptionskip}{-0.5cm}
  \centering
  \includegraphics[scale = 0.46]{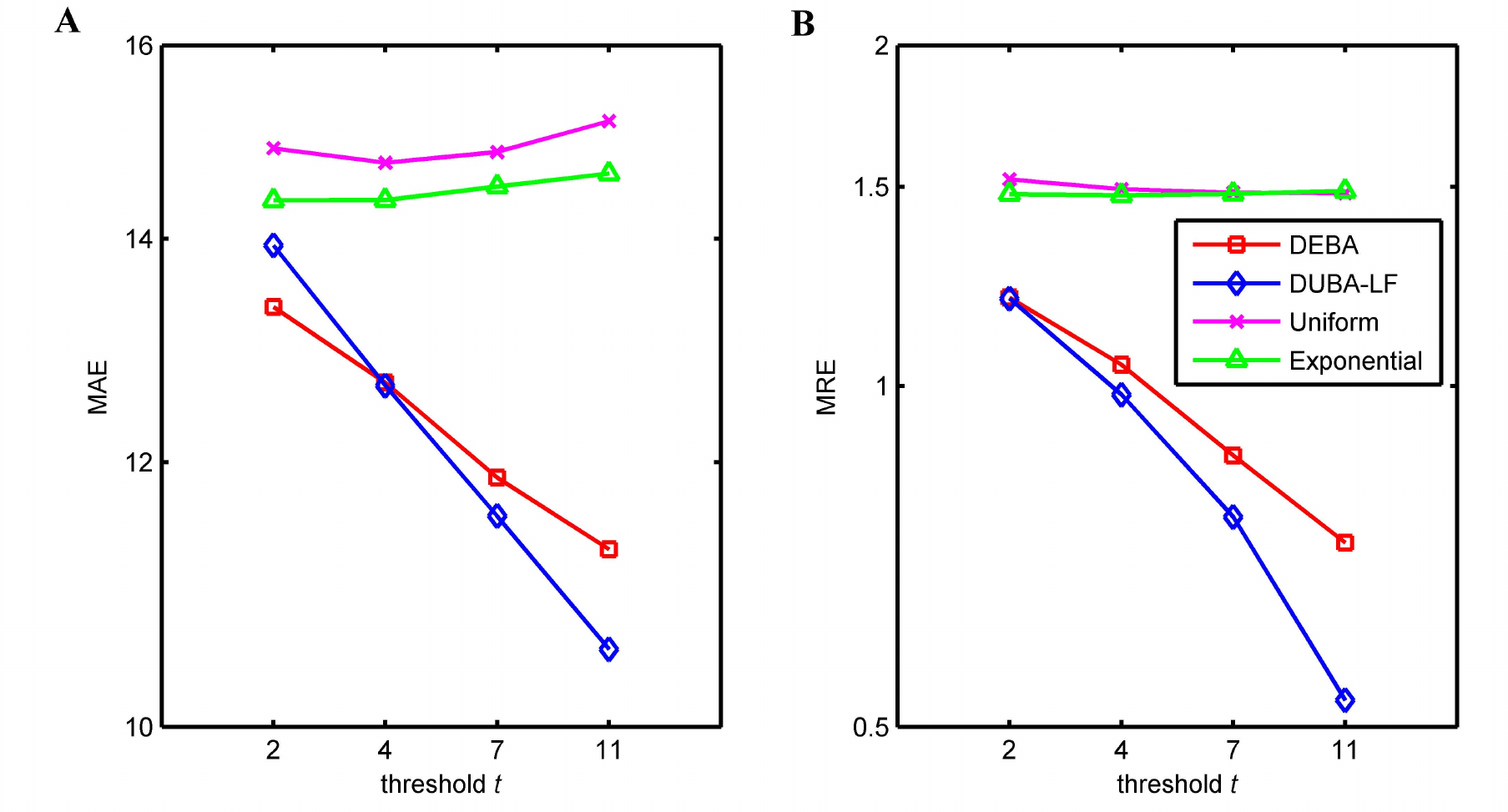}
  \caption{MAE and MRE of each mechanism for the Polblogs dataset, as sample threshold is varied and $c$ is set to 7.}
\end{figure}

Fig.2 plots the MAE and MRE of all schemes for the Polblogs dataset, where we vary the sampling threshold t and set $c$ = 7.  DUBA-LF is the best method in the measurement of MRE but is outperformed by DEBA in MAE for small sampling threshold. For the sampling threshold $t (t\geq4)$, DUBA-LF outperforms \emph{Uniform} mechanism by up to 43.9\% in MAE and 180.4\% in MRE, \emph{Exponential} by up to 38.8\% in in MAE and 181.8\% in MRE, and DEBA by up to 7.1\% in MAE and 37.8\% in MRE. The increasing the sample threshold has an effect of decreasing of MAR and MRE for DUBA-LF and DEBA mechanisms but the effect for \emph{Uniform} and \emph{Exponential} mechanism is less.

\begin{figure}[h]
  \setlength{\abovecaptionskip}{0.cm}
  \setlength{\belowcaptionskip}{-0.5cm}
  \centering
  \includegraphics[scale = 0.46]{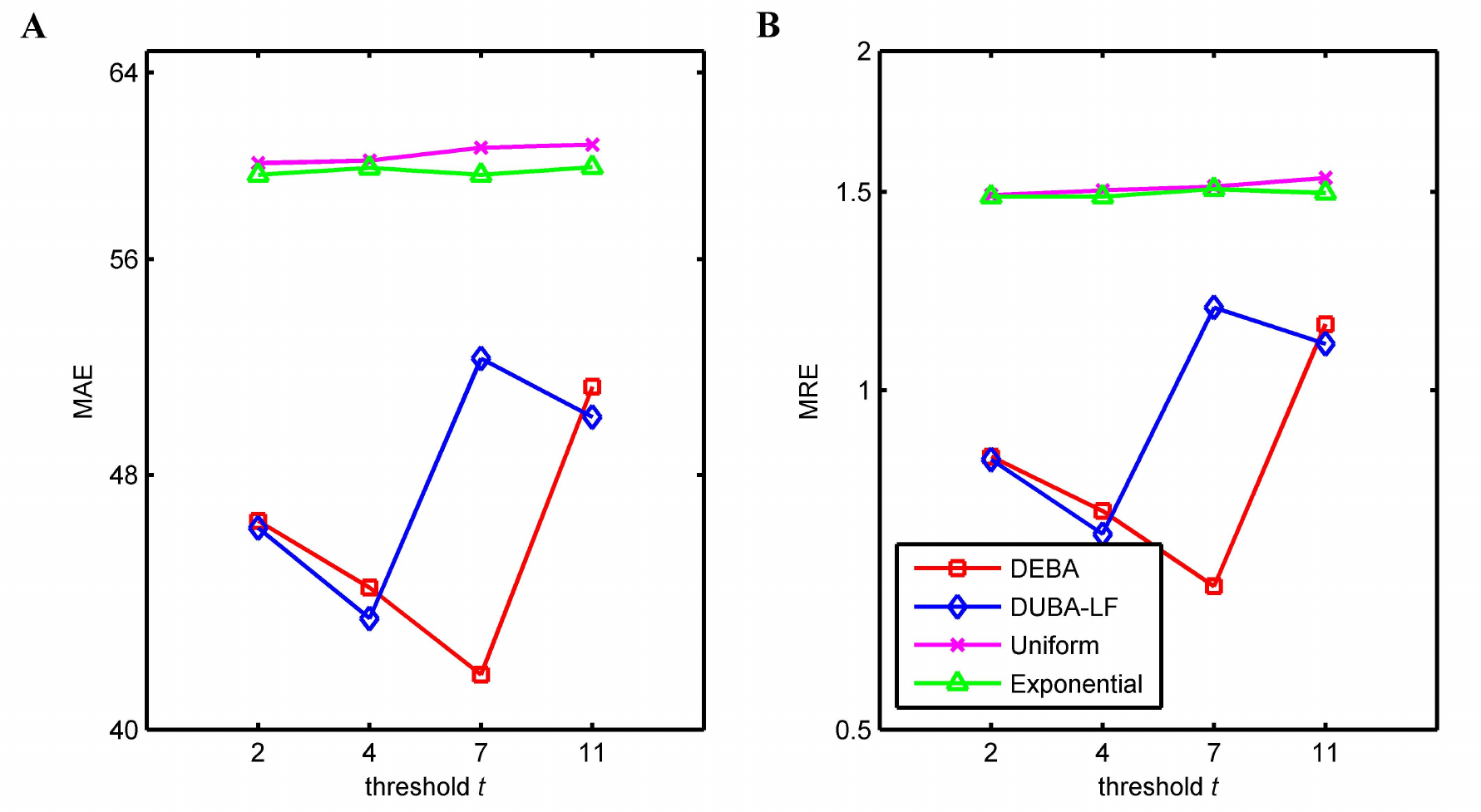}
  \caption{MAE and MRE of each mechanism for the Facebook dataset, as sample threshold is varied and $c$ is set to 4.}
\end{figure}

Fig.3 shows the MAE and MRE of all schemes for the Facebook dataset, where we vary the sampling threshold $t$ and set $c = 4$. DEBA seems to outperform the other methods in this setting. And DEBA outperforms \emph{Uniform} mechanism by up to 45.8\% in MAE and 126.3\% in MRE, \emph{Exponential} by up to 42.9\% in in MAE and 125.0\% in MRE, and DUBA-LF by up to 25.4\% in MAE and 76.7\% in MRE. Similar to Fig.2, increasing the sampling threshold has a less evident effect for both \emph{Uniform} and \emph{Exponential} mechanisms. We can also conclude that increasing the sampling thresholds causes decreasing of MAR and MRE for DUBA-LF and DEBA for small threshold $t$ while increasing of MAR and MRE for larger $t$.
\begin{figure}[h]
  \setlength{\abovecaptionskip}{0.cm}
  \setlength{\belowcaptionskip}{-0.5cm}
  \centering
  \includegraphics[scale = 0.46]{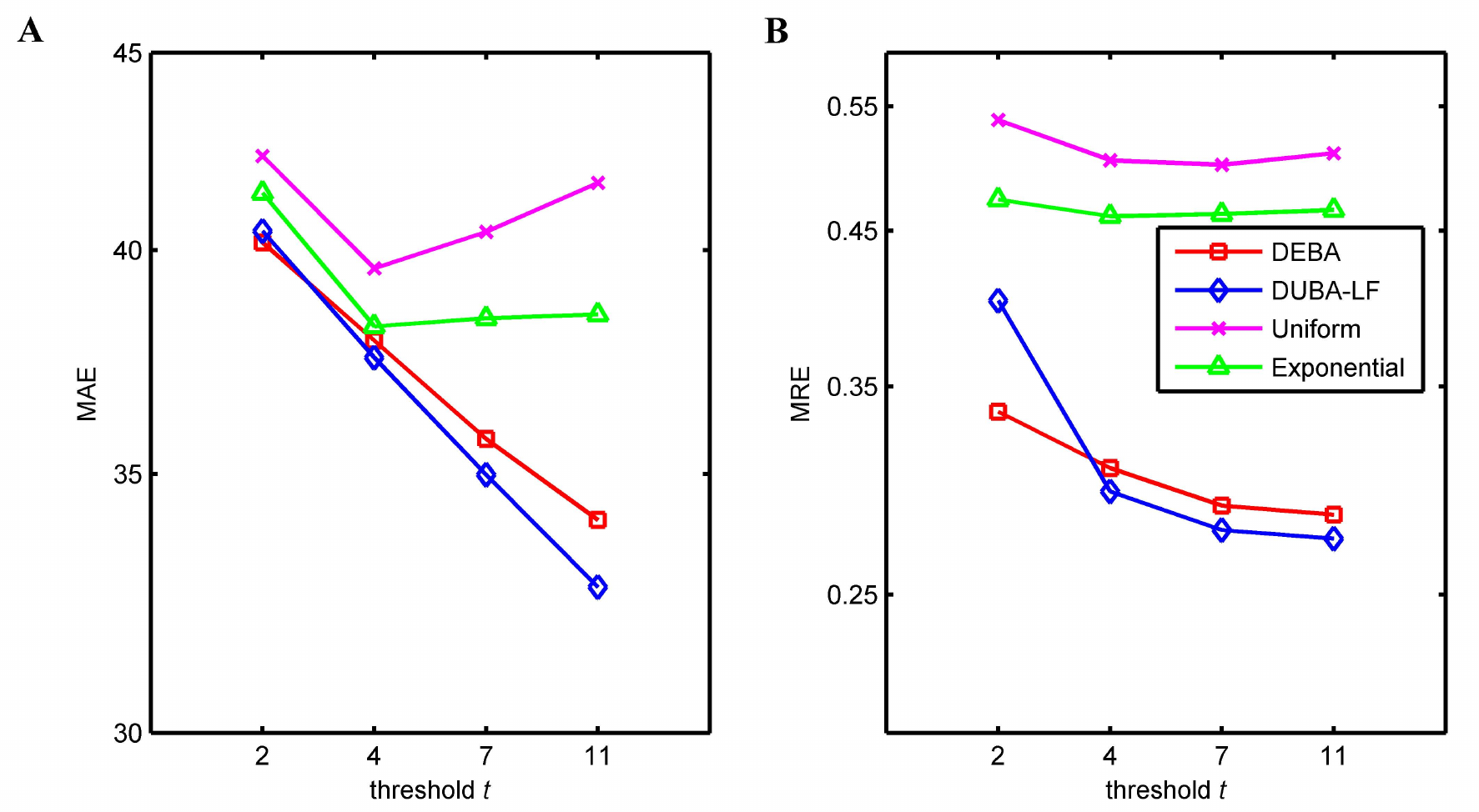}
  \caption{MAE and MRE of each mechanism for the Facebook dataset, as sample threshold is varied and $c$ is set to 7.}
\end{figure}

The MAE and MRE of all schemes for the Facebook dataset is showed in Fig.4, and here we set $c = 4$ and vary the sampling threshold $t$. DUBA-LF seems to outperform the other methods in this setting. It outperforms \emph{Uniform} mechanism by up to 27.2\% in MAE and 86.3\% in MRE, \emph{Exponential} by up to 17.7\% in in MAE and 70.0\% in MRE, and DEBA by up to 4.1\% in MAE and 3.9\% in MRE. Similar to Fig.1, increasing the sampling thresholds causes decreasing of MAR and MRE for DUBA-LF and DEBA. However, increasing the sample threshold has an evident effect for decreasing of MAR and MRE for both \emph{Uniform} and \emph{Exponential} for small threshold $t$ while increasing of MAR and MRE for larger $t$.

\begin{figure}[h]
  \setlength{\abovecaptionskip}{0.cm}
  \setlength{\belowcaptionskip}{-0.5cm}
  \centering
  \includegraphics[scale = 0.46]{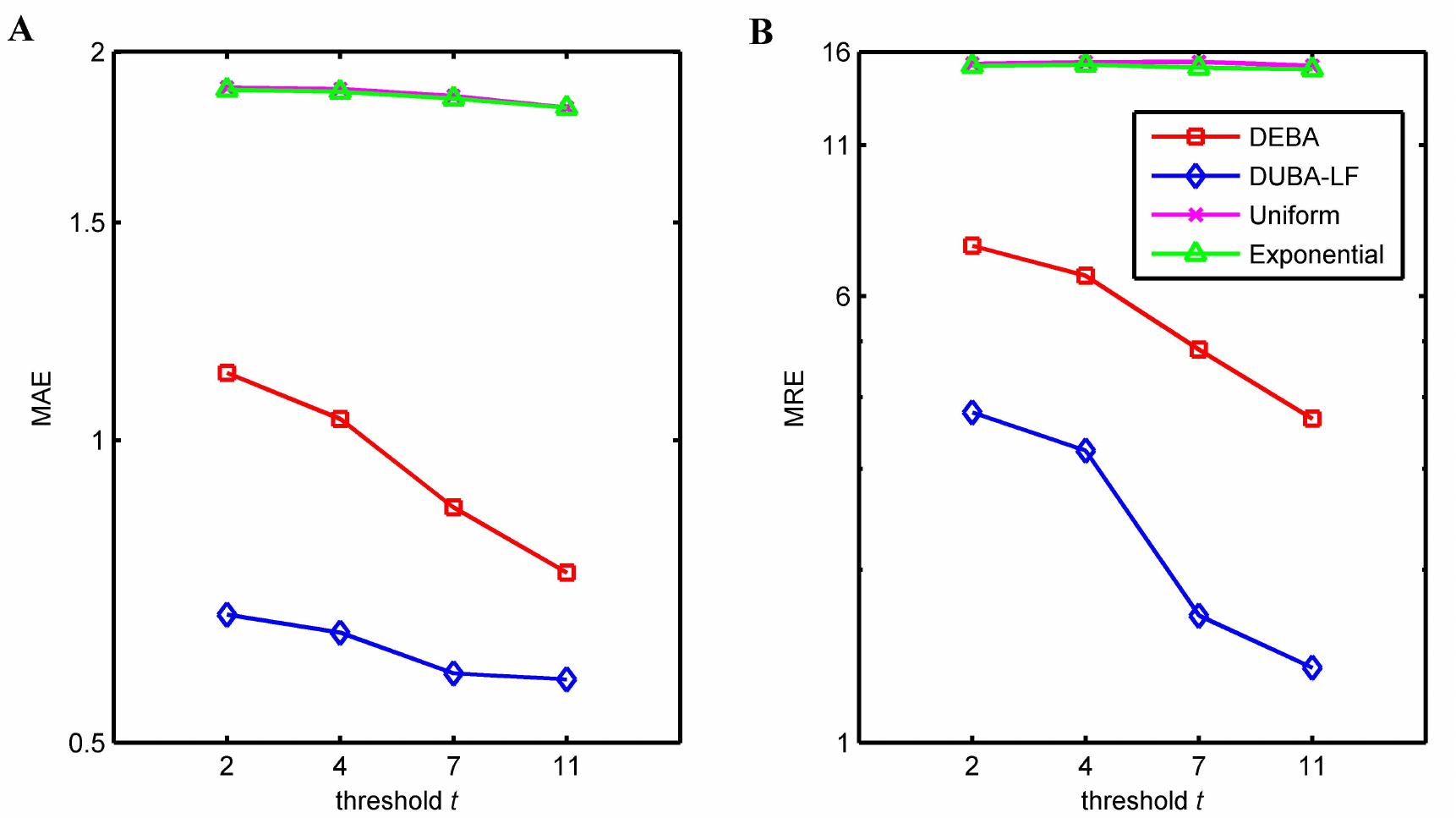}
  \caption{MAE and MRE of each mechanism for the CondMat dataset, as sample threshold is varied and $c$ is set to 4.}
\end{figure}

Fig.5 plots the MAE and MRE of all schemes for the CondMat dataset, where the sample threshold $t$ is varied and $c$ is set to 4. DUBA-LF is the best method in this setting. And it outperforms \emph{Uniform} mechanism by up to 107.9\% in MAE and one order of magnitude in MRE, \emph{Exponential} by up to 107.2\% in in MAE and also one order of magnitude in MRE, and DEBA by up to 35.8\% in MAE and 190.7\% in MRE. The results also indicate that increasing the sampling threshold has an effect of decreasing of MAR and MRE for DUBA-LF, DEBA and \emph{Uniform} mechanisms but the effect for \emph{Exponential} mechanism is not evident.

\begin{figure}[h]
  \setlength{\abovecaptionskip}{0.cm}
  \setlength{\belowcaptionskip}{-0.5cm}
  \centering
  \includegraphics[scale = 0.46]{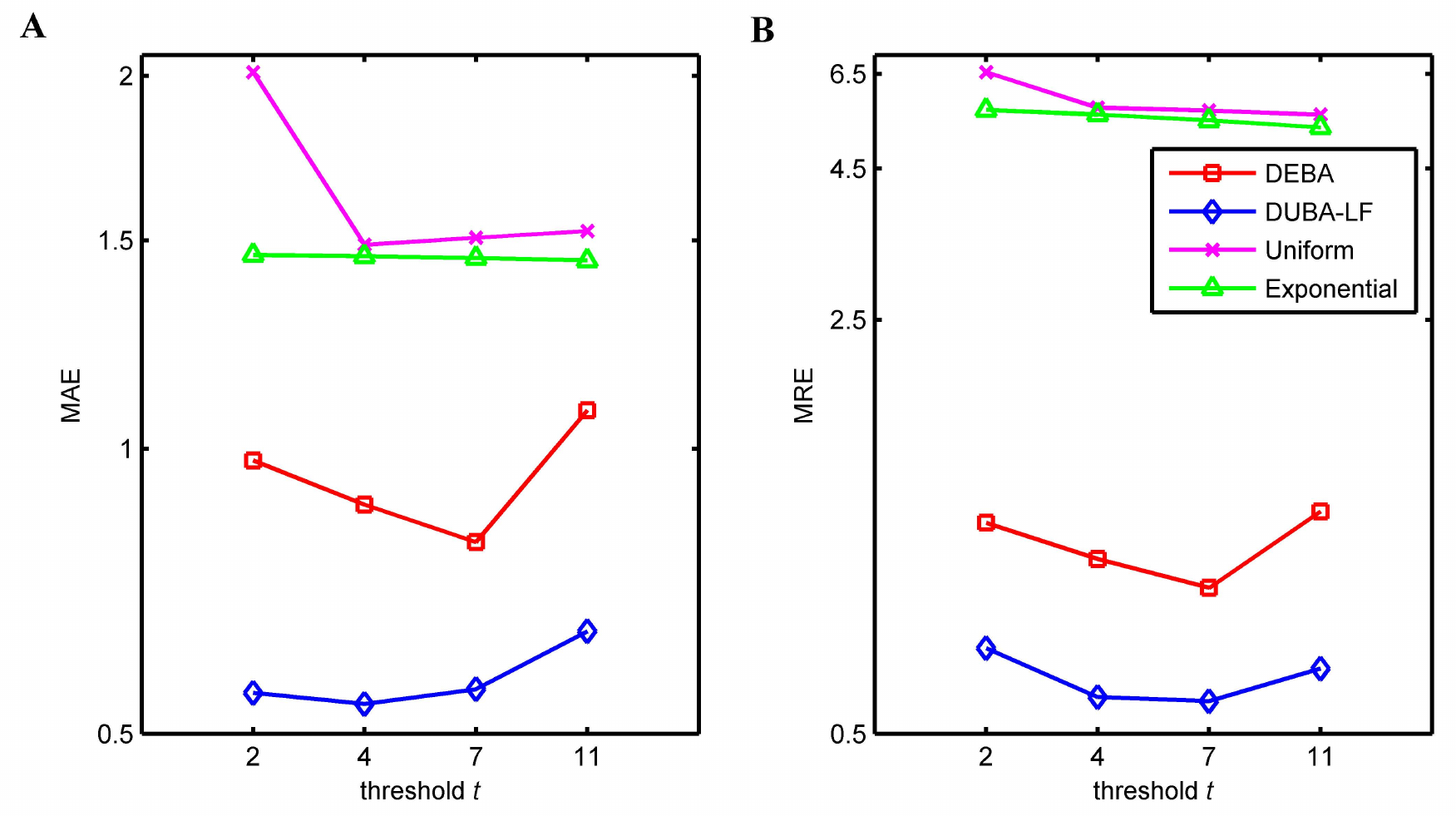}
  \caption{MAE and MRE of each mechanism for the CondMat dataset, as sample threshold is varied and $c$ is set to 7.}
\end{figure}

Fig.6 shows the MAE and MRE of all schemes for the CondMat dataset, where the sample threshold $t$ is varied and $c$ is set to 7. DUBA-LF is shown to be the best method in this setting. And it outperforms \emph{Uniform} mechanism by up to  120.5\% in MAE and  894.5\%  in MRE, \emph{Exponential} by up to 76.9\% in in MAE and  862.7\% in MRE, and DEBA by up to 34.4\% in MAE and 84.3\% in MRE. It is indicated that increasing the sampling threshold has a little effect of decreasing of MAR and MRE for \emph{Uniform} and \emph{Exponential} mechanisms. We can also conclude that increasing the sampling thresholds causes decreasing of MAR and MRE for DUBA-LF and DEBA for small threshold $t$ while increasing of MAR and MRE for larger $t$.

\section{Conclusion}

The number of CFPs is one of the most important properties for a public users labeled graph. In order to release the number of CFPs in the context of personalized privacy preferences, we proposed two schemes (DEBA and DUBA-LF) to achieve personal differential privacy. Both DEBA and DUBA-LF use the distance-based budget absorption mechanism to improve the publication utility while DUBA-LF also employs ladder function to reduce the introduced noise. We formally prove that the proposed DEBA and DUBA-LF schemes are $\mathcal{P}$-PDP and we conduct thorough experimentation with real datasets, which demonstrated the superiority and the practicality of our proposed schemes.

\end{document}